# Beyond the Linear Damping Model for Mechanical Harmonic Oscillators

**Randall D. Peters**

**Department of Physics**
**1400 Coleman Ave.**
**Mercer University**
**Macon, Georgia 31207**

**19 August 2004**

## Abstract

The steady state motion of a folded pendulum has been studied using frequencies of drive that are mainly below the natural (resonance) frequency of the instrument. Although the free-decay of this mechanical oscillator appears textbook exponential, the steady state behavior of the instrument for sub-resonance drive can be remarkably complex. Although the response cannot be explained by linear damping models, the general features can be understood with the nonlinear, modified Coulomb damping model developed by the author.

**Introduction**

Previous studies of mechanical oscillators, particularly those involving long-period pendula, have provided evidence for the need to be cautious when using linear models to describe harmonic oscillator damping (see, for example [1]). For example, the naive use of viscous damping to describe the torsion pendulum has been cause for serious disagreement among experimenters trying to improve our knowledge of the Newtonian gravitational constant G [2]. Internal friction is more important than fluid damping for long-period mechanical oscillators. `Hysteretic' damping, the single-most important dissipation mechanism for these harmonic oscillators, has been long used by engineers but not by physicists. Yet it was the work of physicists Kimball and Lovell [3] that introduced the world to this universal form of internal friction. The `signature' of hysteretic damping becomes evident in a plot of quality factor Q of the oscillator against frequency f. Unlike the viscous damping model where $Q \propto f$, hysteretic damping predicts $Q \propto f^2$. This behavior has been observed with a variety of different mechanical oscillators, perhaps first in a LaCoste spring vertical seismometer by Gunar Streckeisen [4]. It should be noted that the quadratic dependence reduces to a `non-dependence' when the Q of the instrument drops below the critical damping value of 1/2. Thus mechanical spectroscopy measurements in which there is no frequency dependence to the internal friction (because the system has too little inertia to oscillate) is consistent with universal (hysteretic) friction as described by Kimball and Lovell. To designate internal friction for this case by $Q^{-1} \propto f^0$ is a misleading though common nomenclature.

In Fig. 1 it is seen that the Q of the folded pendulum of the present study also displays the universal feature.

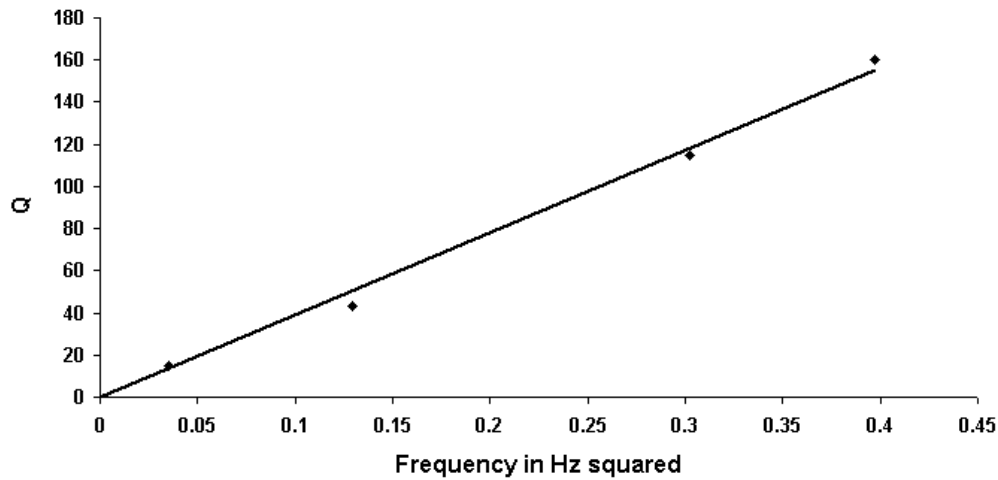

**Figure 1.** Frequency variation of the quality factor of the folded pendulum of the present study.

**Differences-Linear vs Nonlinear Models**

# Beyond the Linear Damping Model for Mechanical Harmonic Oscillators

The viscous damping (linear) model is widely known; and its sometimes serious limitations have been mostly overlooked. Being one of the best known differential equations in all of physics, the equation of motion for the harmonic oscillator with viscous damping is expressible as (free-decay, no drive):

$$\ddot{x} + 2\beta \dot{x} + \omega_0^2 x = 0 \qquad (1)$$

where $\beta = \omega/(2Q)$ is the damping coefficient, also sometimes erroneously labeled `constant'. In [5] it is shown from an experiment involving fluid damping - that the coefficient depends not only on fluid viscosity but also on fluid density and it is also frequency dependent. The only thing truly `simple' about harmonic oscillators, for which Eq. 1 has been `sacrosanct'- is the overly-idealized mathematics of Eq.(1).

The modified Coulomb (nonlinear) damping model developed by the author [6] is for the present work (exponential free-decay) expressible as

$$\ddot{x} + \frac{\pi\omega}{4Q}[\omega^2 x^2 + \dot{x}^2]^{1/2} \operatorname{sgn}(\dot{x}) + \omega^2 x = 0 \qquad (2)$$

The friction force is seen to be always opposite in direction to the velocity (through the function sgn); however, its magnitude is proportional to the square root of the energy of oscillation. Notice that there is no subscript zero on the frequency in Eq.(2), since unlike the viscous damping model Eq.(2) predicts no `damping redshift'. Even the damping redshift predicted by Eq.(1) is so small as to defy measurement; thus the great attention to it in textbooks is mostly unwarranted (c.f. ref. [1 and 7]).

A comment about quality factor is warranted. It is defined by $Q = 2\pi E/\Delta E$ where E is the energy of oscillation and $\Delta E$ is the energy loss (work against friction) over one cycle of the motion. The damping term is presented in the literature in a variety of ways; yet every case known to the author can be expressed in the canonical form $(\omega/Q) \times [v]$, where $[v] = dx/dt$ **only** in the case of viscous damping. For models such as Eq.(2), where the damping term is responsible for the nonlinear properties of the equation, $[v] = (\pi/4)[\omega^2 x^2 + (dx/dt)^2]^{1/2} \operatorname{sgn}(dx/dt)$, and the decay is exponential. For Coulomb (sliding) friction $[v] = \text{const} \times \operatorname{sgn}(dx/dt)$, and a fit to the turning points of the decay is not an exponential but rather a straight line. It is the postulate of the present paper that the one thing common to the damping force and the velocity in real mechanical oscillators, dominated by internal friction, is the direction of the vectors. As specified in Eq.(2) the damping is always opposite the velocity in direction. It should be noted however, that the direction of the damping force can be the same as the velocity (negative damping) for oscillators, that have been put in a far-from-equilibrium state, such as the laser driven pendulum of ref. [8]. Moreover, later in this article is presented evidence that even systems close to thermodynamic equilibrium can still exhibit negative damping during a fraction of the oscillator's cycle.

A free-decay comparison of Equations (1) and (2) is provided in Fig. 2, which was generated for each case by numerically integrating each of the differential equations.

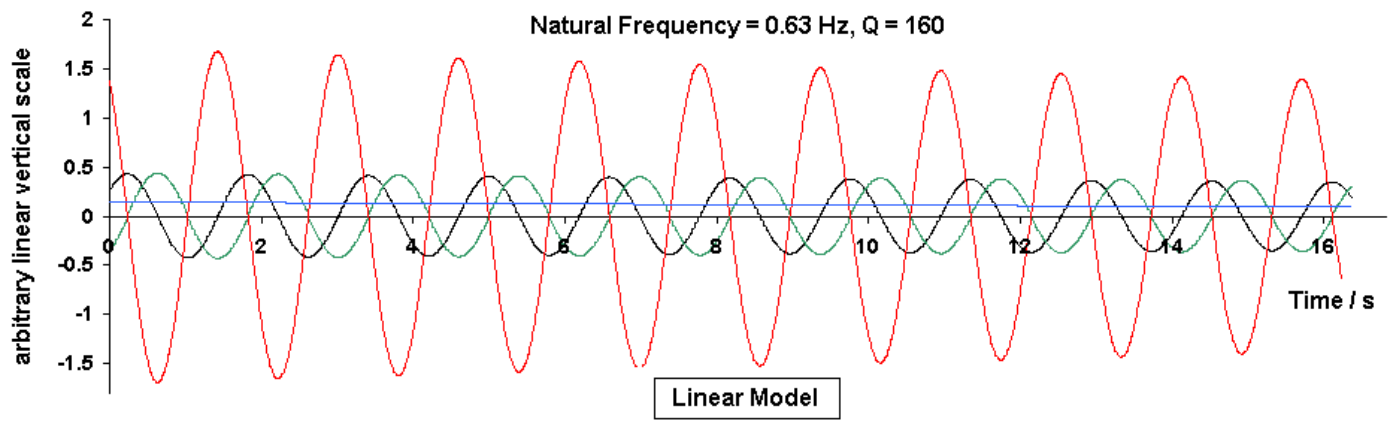

Linear Model

LEGEND: Position (black), Velocity (red), 0.1 x Energy (blue), 10 x friction force (green)

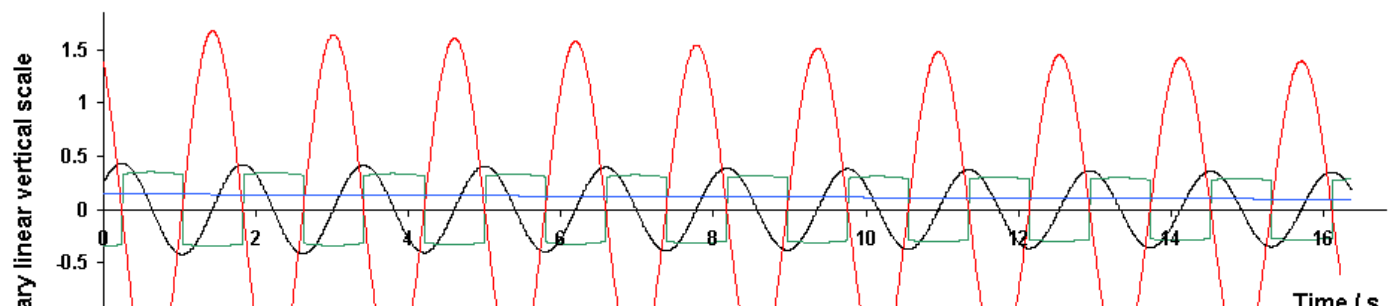

Beyond the Linear Damping Model for Mechanical Harmonic Oscillators

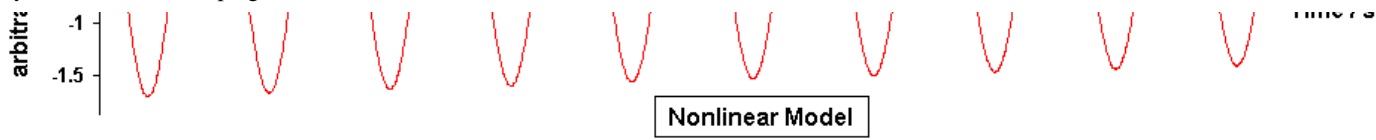

**Figure 2.** Theory generated free-decay records showing the difference (green traces, assuming unit mass) in the effective internal friction force, between the linear and nonlinear damping models.

It can be readily shown, from Fourier series considerations, that the fundamental alone is responsible for the Q of the oscillator, no matter the form of the periodic friction force. This explains why a damping term proportional to velocity is such a mathematically useful approximation. Notice that the nonlinear decay curve of Fig. 2 is one in which the friction force (magnified 10 times for the plot) is nearly an exponentially decaying square wave. Thus one sees the reason for the multiplier on the square root term in Eq.(2); i.e., the fundamental Fourier component is $4/\pi$ times the amplitude of the square wave.

Subtle evidence for nonlinear damping is uncovered when one does a careful comparison of the difference (residuals) between an actual free decay and a waveform generated from Eq.(1). In an ideal world, a spectral analysis [fast Fourier transform (FFT)] of the residuals should serve to identify if the damping is nonlinear. In actuality, the presence of mechanical `noises' do not provide indisputable evidence in support of Eq.(2). The nonlinearity is masked by internal friction complexities originating at the mesoscale of the polycrystalline metals from which the oscillator is fabricated [7]. On the other hand, the present study shows that external drive of the mechanical oscillator provides a clear means to identify damping nonlinearity.

**Driven system**

When an oscillator described by a linear damping model is driven, motion at the natural frequency of the oscillator decays toward zero as a transient. This occurs as the oscillator entrains with the source. Entrainment is realized regardless of whether the drive frequency is above or below the value for resonance. We will see that the nonlinear model of Eq.(2) results in significantly different behavior-that some sub-resonance drive frequencies result in significant periodic motion at the resonance frequency of the oscillator, after steady state has been reached.

**Sub-resonance Drive**

Consider drive frequencies below the resonance (natural) frequency of the oscillator (sub-resonance drive). In the curves that follow, only nonlinear damping is considered, since the linear model disallows the phenomena here presented for the first time. Shown in Fig. 3 are two examples of sub-resonance drive. In each case numerical integration continued until steady state was achieved, after which the graph was generated.

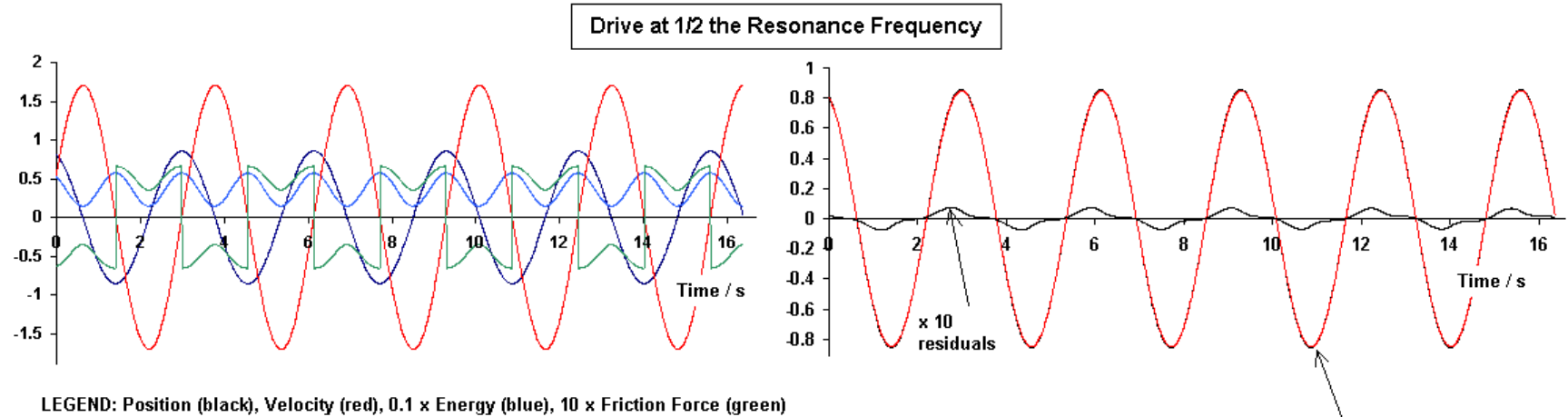

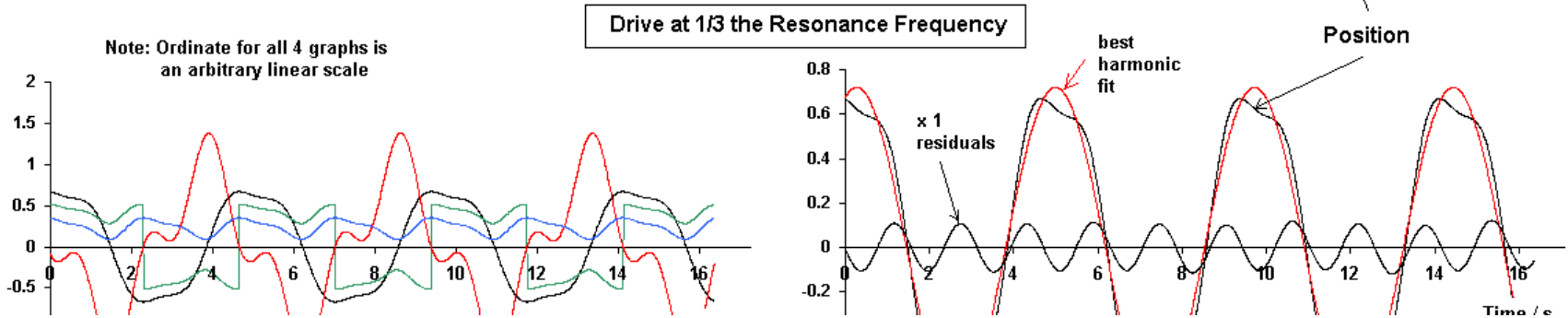

Beyond the Linear Damping Model for Mechanical Harmonic Oscillators

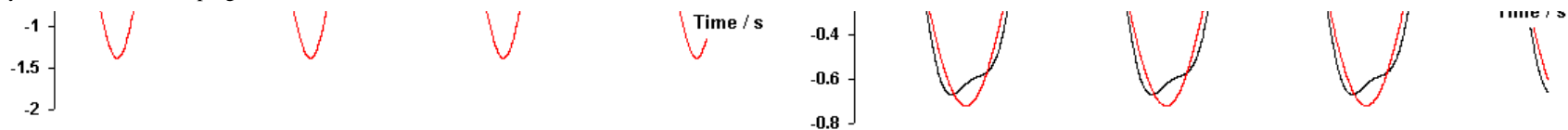

**Figure 3.** Two examples of steady state response to a sub-resonance drive, generated using Eq.(2), but with a drive term on the right-side of the equation.

In the model software the resonance frequency was set at 0.63 Hz and the Q at 160. For the top pair of graphs the drive frequency was set at (0.63 Hz)/2, and for the bottom pair at (0.63 Hz)/3. The left-side plot for each case involves the same variables as in Fig. 2. The right-side plot for each case shows the difference (residuals) between position data and a harmonic best fit to the position. Unlike free-decay (and also drive frequency above resonance), the 1/3 drive case shows a significant component of the motion having a frequency of 0.63 Hz. In other words, the nonlinear friction force is responsible for a 3rd harmonic response to the drive-equivalent to a persistent (steady-state rather than transient) motion at the natural frequency of the oscillator.

**Experimental Setup**

Shown in Fig. 4 is a photograph of the folded pendulum. There are four axis `hinges', three of which are glass microscope slides with one sharp (90-degree) edge resting on a ceramic flat. Mounted with epoxy at a 45 degree angle to a member, the edge of the slide rests against a flat of the complementary member. The three hinges of this type are labeled in the photograph as `top', `lower', and `post lower'. The 4th hinge (post upper) is a tungsten carbide tooth `point' hinge, taken from a circular saw blade, and epoxied to the top of the brass post. The inclined member which supports the weight of the pendulum at the top hinge is a hollow tube of fuzed quartz, epoxied at the bottom to the aluminum bar through which a hole had been drilled at an angle.

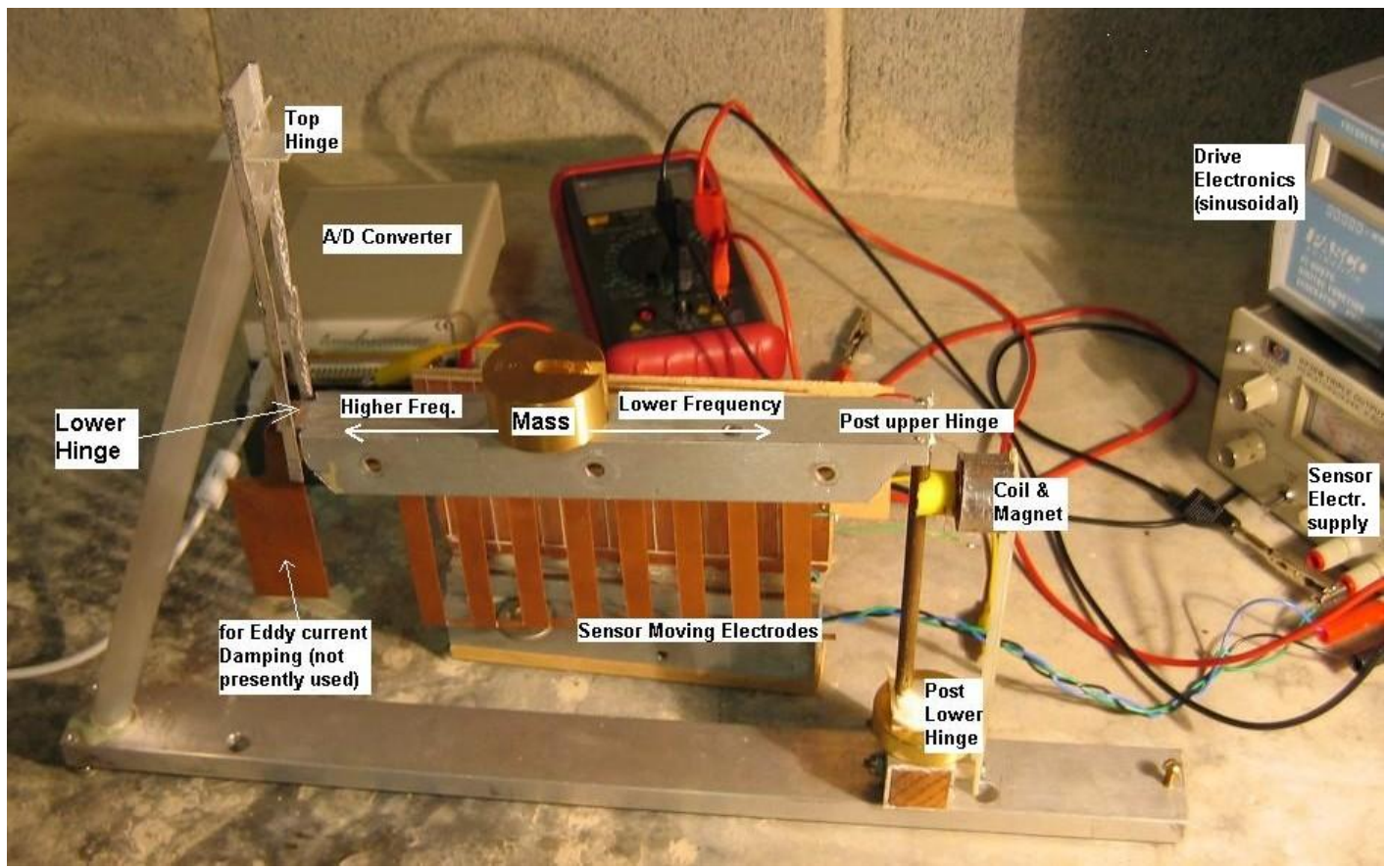

**Figure 4.** Photograph of the folded pendulum with the stationary electrodes of the sensor removed from their operating position.

To adjust the frequency of oscillation, the mass is moved on its supporting boom-toward the post hinge (inverted pendulum side) to lower the frequency and in the opposite direction (toward the `regular' pendulum side) to increase the frequency. Drive is provided by a sinusoidal constant amplitude voltage source (Pasco PI 9587C), the output of which is connected through a 3K resistor to the coil of the coil/magnet (Faraday law) drive system. The coil is fixed to the post by means of the yellow plastic member shown.

Since spectral analysis is key to the conclusions reached, it was imperative that the Pasco oscillator be free from harmonic distortions. Spectral analysis of the drive voltage records using the FFT showed the oscillator's output to be nearly perfectly sinusoidal, with distortion components (2nd harmonic being the largest) at least 60 dB (factor of 1000) below the fundamental.

This instrument had been previously functioning as a horizontal seismometer; in this mode the copper plate on the left of the boom was positioned between strong, rare earth permanent magnets to provide eddy-current damping. For the present experiment the damping magnets were removed.

The sensor for this work is an eight-element array form of the symmetric differential capacitive (SDC) sensor patented by the author [9]. The sensor moving electrodes, glued to the boom, execute nearly linear motion over a range of several mm's. The eight arms of this array are visible in the photograph, but the associated static electrodes that straddle these moving electrodes have for this photograph been removed from their operating position.

The power supply employed by the sensor electronics is partially visible; it is a Hewlett Packard HP6236B. For the voltage settings presently used (+/- 9 V), the calibration constant of the sensor was measured to be 20,000 V/m.

**Experimental Results & Comparison with Theory**

Shown in Fig. 5 is a free-decay record taken with the instrument operating at 0.63 Hz. Similar decay records were collected for each of the other three points noted in Fig. 1. As the period of the instrument lengthens, the decay becomes increasingly less perfect than Fig. 5, for a variety of factors related to the increased sensitivity at lower frequency (the sensitivity being proportional to the square of the period). In all the cases presently reported, the free-decay records were found to be essentially exponential.

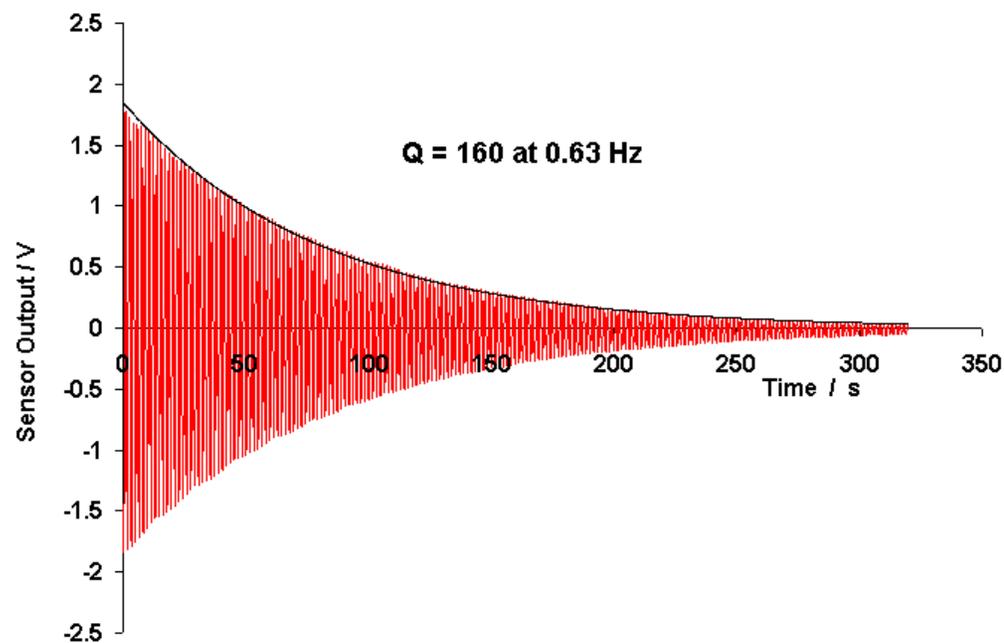

**Figure 5.** Sample Free-Decay record of the folded pendulum

The plot was generated by importing the data to Excel after first saving the A/D stored record in csv (*.dat) format. The analog to digital converter used for the experiments is a Dataq DI-700. For the sensor constant of 20,000 V/m it is seen from Fig. 5 that the initial amplitude of the motion was 90 μm.. Because of the nearly perfect exponential decay of Fig. 5 (with black-line fit shown), one might be tempted to declare that the system is completely linear. We will see from other data which follows that the conclusion of linear damping is false.

**Sub-Resonance Drive Results**

Shown in Fig. 6 are the experimental results of driving the oscillator at 1/3 its resonance frequency.

Beyond the Linear Damping Model for Mechanical Harmonic Oscillators

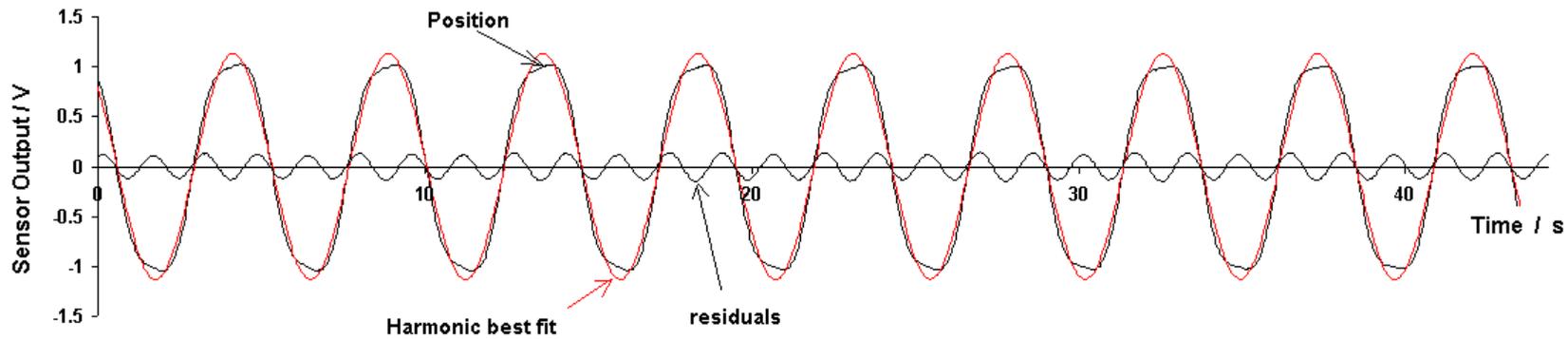

**Figure 6.** Experimental evidence for nonlinear damping in the response of the folded pendulum to a 1/3 sub-resonance drive at steady state.

It is seen that the oscillator responds to a 1/3 sub-resonance drive essentially as predicted by the nonlinear damping model [Eq.(2)] that was used to generate the lower right graph of Fig. 3. The driven-case response is slightly more noisy than expected on the basis of the free decay record, as illustrated by the spectra of Fig. 7. These plots were generated with the first 2048 points of each record, using Excel's Fourier analysis routine.

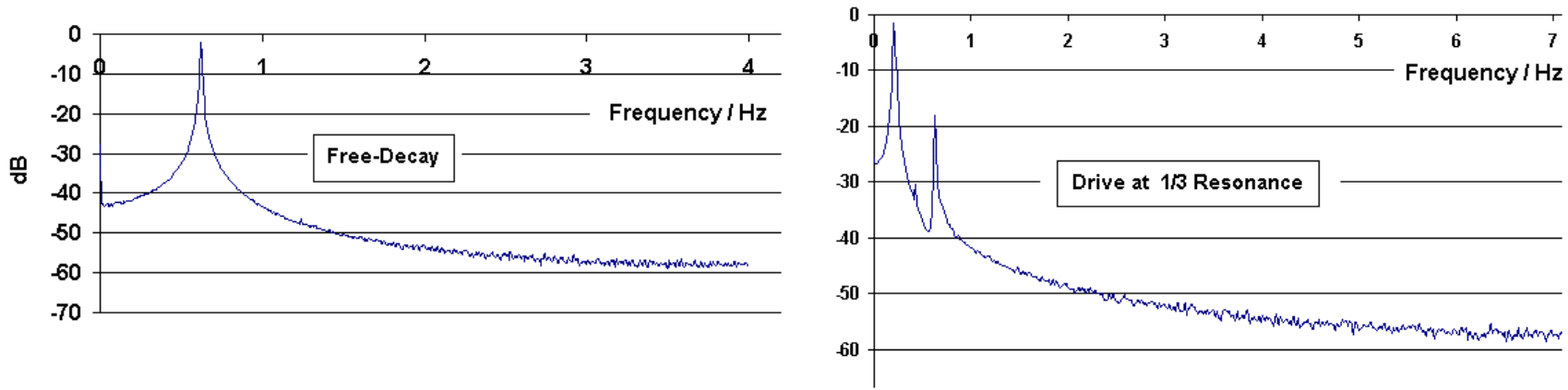

**Figure 7.** Comparison of experimental Spectra, free-decay and driven.

Although agreement between Eq.(2) and experiment are reasonable for the case of drive at 1/3 resonance, the same is not true for drive at 1/2 resonance, as seen from Fig. 8. The upper right graph of Fig. 3 predicts that there should be no significant retention of motion at the frequency of resonance when the pendulum is driven at 1/2 the resonance frequency.

Beyond the Linear Damping Model for Mechanical Harmonic Oscillators

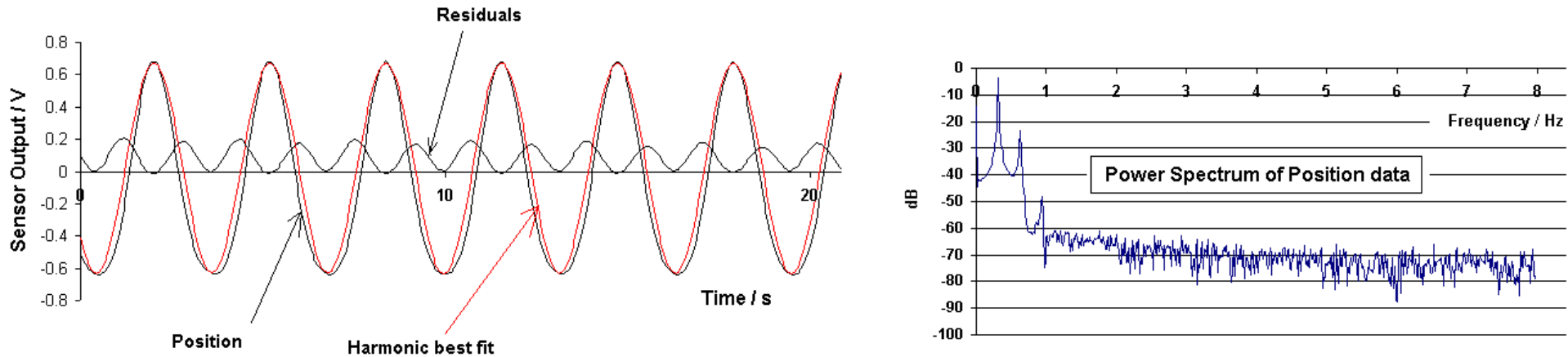

**Figure 8.** Experimental results for drive at 1/2 the resonance frequency of 0.63 Hz.

The nonlinear damping model is easily modified to accommodate anisotropic friction, a change which provides agreement between theory and experiment. Such a modification was made to the code before generating Fig. 9.

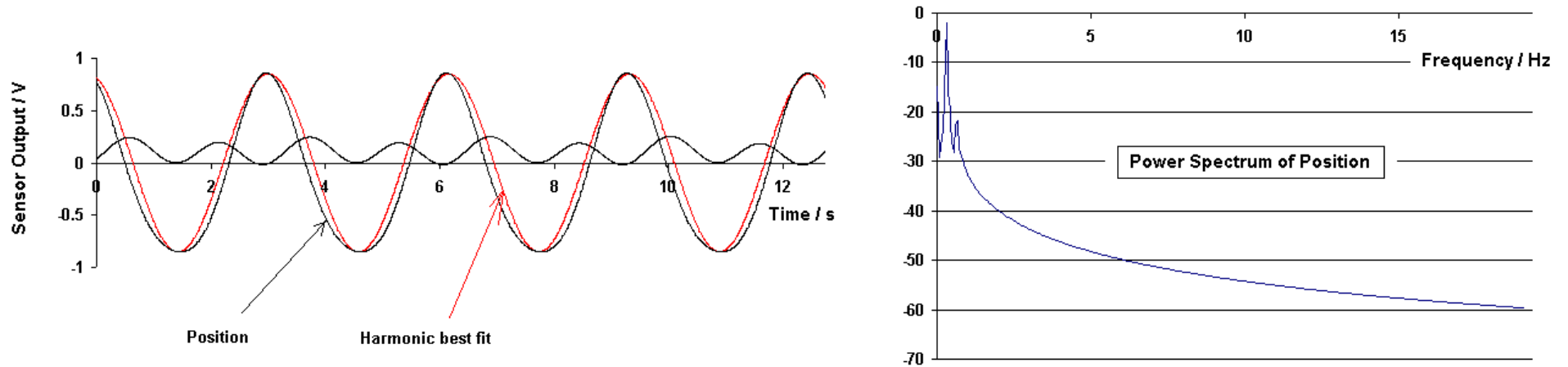

**Figure 9.** Model predicted response to drive at 1/2 the resonance frequency and using anisotropic friction.

**Anisotropic Friction**

To understand the case presented in Fig. 9, let the friction (square root term) in Eq. (2) be represented by the number pair 1,-1. This means the friction force is opposite the velocity at all times and has the same magnitude during both half-cycles of the motion. For the graph of Fig. 9, the friction was set instead at 3,1. In other words, for one set of half-cycles the friction force is in the same rather than opposite direction of the velocity (so-called negative damping) and has the same magnitude as the isotropic case; but for the other set of half-cycles the damping is positive but three times greater than the isotropic case. The difference of 3 -1 = 2 is cause for the same Q = 160. A comparison of Figures 8 and 9 shows that the anisotropic nonlinear damping model is in reasonable agreement with the experimental case shown. One point of difference is worthy of special note. The experimental spectrum shows a line having frequency 0.31+0.63= 0.94 Hz; a characteristic that is possibly explained by elastic nonlinearity (not treated in the present publication).

# Beyond the Linear Damping Model for Mechanical Harmonic Oscillators

**Frequency dependence of the Anisotropy**

Whereas negative damping is required as part of the anisotropy to explain Fig. 8, it may not be required at every frequency of operation. For example, shown in Fig. 10 is the steady state response for the 0.15 Hz eigenfrequency case.

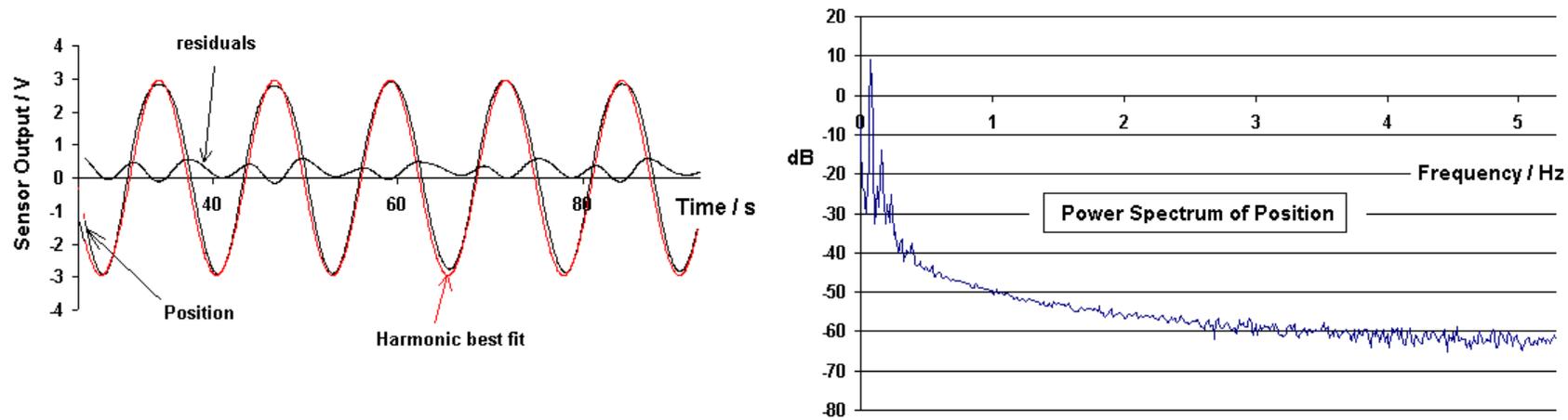

**Figure 10.** Experimental results for drive at 1/2 the resonance frequency of 0.15 Hz.

Notice that the amount of distortion has decreased as compared to Fig. 8, and there has also been a reversal in the parity of the anisotropy. In Fig. 8 the sharper turning points of the motion are on top, whereas in Fig. 10 they are on the bottom. The change in parity is probably the result of inhomogenities in the metal structures of the pendulum and the repositioning of components as part of lowering the frequency of oscillation. Whether the magnitude of distortion generally decreases with frequency is unknown, since too few frequencies have been considered in relationship to this matter. Since the energy of oscillation decreases at lower frequencies, it is reasonable to think that the distortion might also decrease.

**More general Comparison of Theory and Experiment**

Perhaps the best single graph in support of the nonlinear damping theory is that of Fig. 11. Here the `retention' of the eigenfrequency has been plotted as a function of sub-resonance drive frequency. A 20 dB change in the ordinate scale corresponds to a ten-fold change in voltage, and thus amplitude of the oscillator's displacement.

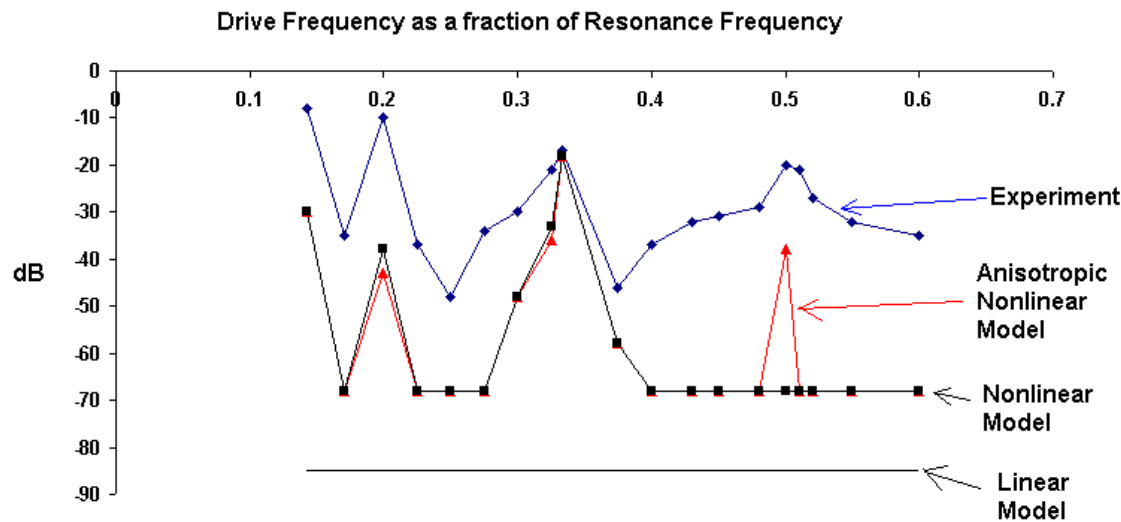

Beyond the Linear Damping Model for Mechanical Harmonic Oscillators

**Figure 11.** Comparison of Experiment and the modified Coulomb nonlinear Damping model as a function of drive frequency at steady state.

The salient features of experiment are described by the model. In earlier theoretical graphs the amplitude of the drive was slightly too small for best agreement with experiment. In Fig. 11 it was set for agreement at the 1/3 drive frequency. At all other frequencies, the experimental values are seen to be higher than the theoretical ones because of noise (mainly mechanical). These noises are probably of 1/f character and intimately connected with mesoscale energy jumps that tend to be in the neighborhood (or multiples of) $1.1 \times 10^{-11}$ J. They were first reported in ref. [10].

**Drive Frequency above Resonance**

Although the present experiment focused mainly on drive below resonance, one case of drive above resonance was treated-that of drive frequency equal to twice the resonance frequency. The experimental data of Fig. 12 shows the presence, at steady state, of motion at the natural (resonance) frequency of the pendulum. This response at 0.63 Hz is 25.4 dB (factor of 18.6) below the primary (drive entraining) response at 1.26 Hz. The residuals (small black curve) of the temporal trace of the left figure shows the natural response to be fairly noisy.

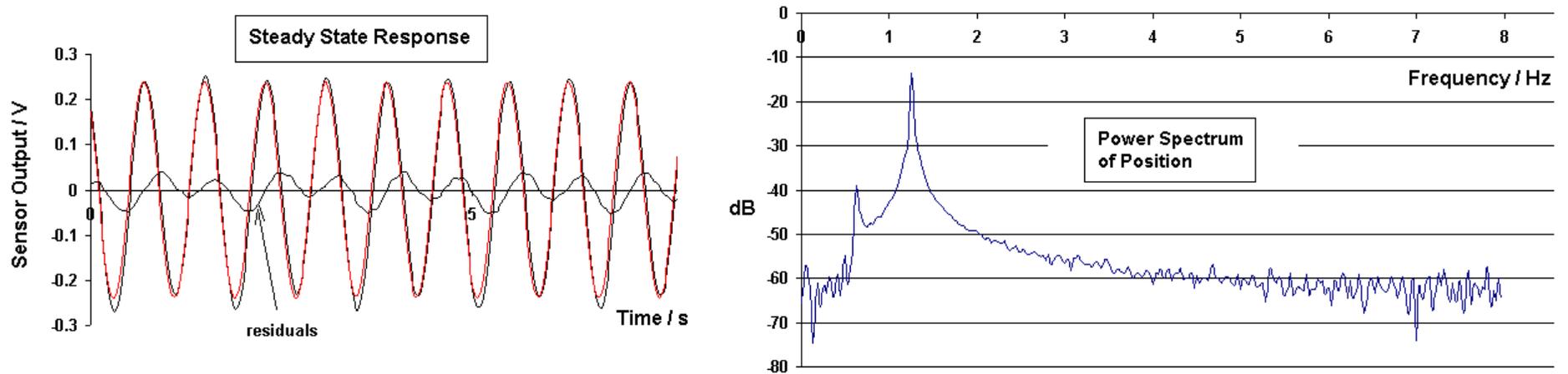

**Figure 12.** Experimental steady-state results for drive frequency at twice the resonance frequency.

The nonlinear model with symmetric friction cannot duplicate the essential (non-noisy) spectrum that is the right-side graph of Fig. 12. However, the anisotropic model with the same 3,1 parameters of earlier comparison does provide reasonable agreement with experiment, as seen from Fig. 13.

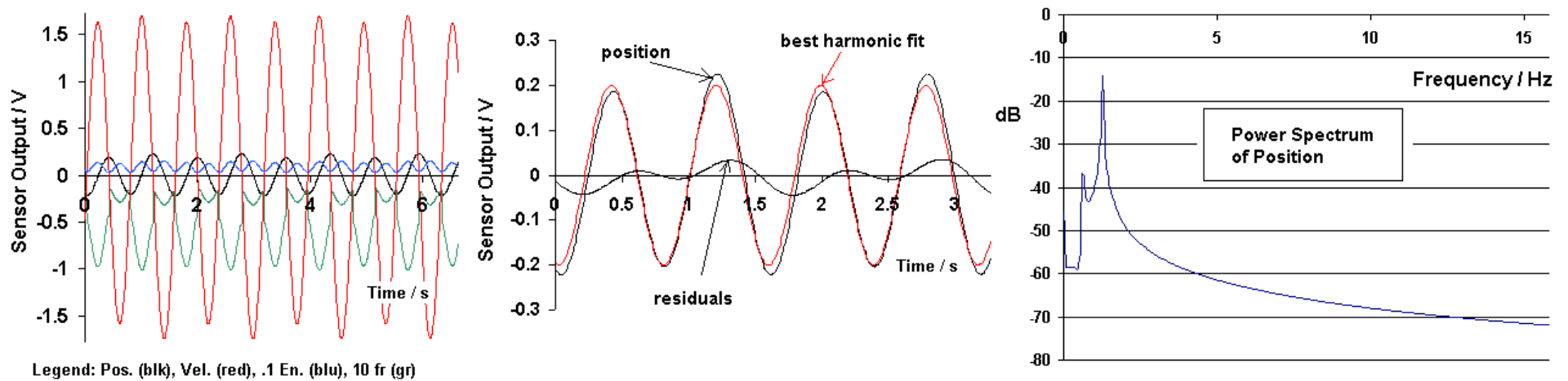

**Figure 13.** Theoretical steady-state anisotropic results for drive frequency at twice the resonance frequency.

**Conclusions**

Response of the pendulum to a harmonic, constant amplitude driving force is not classical and can be surprisingly complex, especially for drive frequencies below resonance. The response at steady state cannot be explained by any linear damping model. Not only is a nonlinear model required, but strong evidence is provided in support for the claim that negative damping occurs for one direction of the velocity in this anisotropic, nonlinear, modified Coulomb damping model.

---